# Leveraging the Mob Mentality: An Experience Report on Mob Programming[*]

Jim Buchan
School of Engineering, Computer & Mathematical Sciences
Auckland University of Technology
New Zealand
jbuchan@aut.ac.nz

Mark Pearl
MYOB NZ Ltd
Newmarket
New Zealand
mark.pearl@myob.com

## ABSTRACT

Mob Programming, or "mobbing", is a relatively new collaborative programming practice being experimented with in different organizational contexts. There are a number of claimed benefits to this way of working, but it is not clear if these are realized in practice and under what circumstances. This paper describes the experience of one team's experiences experimenting with Mob Programming over an 18-month period. The context is programming in a software product organization in the Financial Services sector. The paper details the benefits and challenges observed as well as lessons learned from these experiences. It also reports some early work on understanding others' experiences and perceptions of mobbing through a preliminary international survey of 82 practitioners of Mob Programming. The findings from the case and the survey generally align well, as well as suggesting several fruitful areas for further research into Mob Programming. Practitioners should find this useful to extract learnings to inform their own mobbing experiments and its potential impact on collaborative software development.

## 1 INTRODUCTION

The term Mob Programming was first coined in the Extreme Programming (XP) community in 2003 by Moses Hohman [4] to describe their practice of code refactoring in a group of more than two. The term largely fell into obscurity until Woody Zuill began popularizing it again from 2013 [10]. It was then that Woody began speaking at developer conferences about how his commercial development team of about 8 members was "mobbing" fulltime and the successes they were seeing. Since then many teams all over the world have begun practicing Mob Programming, based on Woody's experiences and explanations. The book by Zuill and Meadows [9] is viewed as the seminal work on Mob programming.

Mob Programming or "mobbing" is when 3 or more people work at a single computer with a large screen to solve code and problems together. Participants in a mob work collaboratively, with one of the team using the single keyboard as the "typist" (sometimes referred to as the "driver"). The typist writes code, mainly at the instruction of the other team members. The others in the mob form the problem-solving team. At regular intervals the typist is swapped, depending on various factors including time at the keyboard, expertise of the individuals and knowledge of the current code base. Woody suggest that there is flexibility about the physical layout and frequency of mobbing to suit each team's particular situation.

Mob Programming is a programming practice that leverages distributed knowledge in real-time and can be seen as taking pair programming to the next level. Distributed knowledge is all the knowledge that a group of people possess and might apply in solving a problem; with Mob Programming the knowledge and the people are brought together in front of a single computer.

An overview of the benefits and challenges of mob programming, based on an analysis of recent literature on Mob Programming is presented in the paper [1]. This paper also identifies some strong themes based on a text analysis of 7 articles and the book by Zuill and Meadows. They find that words such as "learning", "driver", "whole", "retrospective", "defect", "idea", "keyboard", "rotation" and "whole" are emphasized in the conceptualization of mobbing in this literature (p.6). They also identify that more research is needed in the areas of empirical validation, and theoretical rationale, which aligns with the areas of future research proposed in this paper. Given more research like this for mobbing, we could be in the position to apply a framework for research into mobbing similar to the one proposed for pair programming in [3].

This paper shares the experiences of a commercial development team based in New Zealand, who has been practicing Mob Programming for an extended period of time. It complements other experience reports on Mob Programming such as [8] and [2], adding to the body of empirical knowledge in this area.

## 2   BACKGROUND TO THE CASE

The organization provides a cloud-based financial product suite that is developed and maintained in-house. It is a well-established company (over 25 years old) with around 50 development teams. The organization supports self-organizing teams and different teams work in different ways to suit their preferences. This paper focuses on the experience of one development team that has adopted mobbing as one of its software development practices. The Development Lead came to the organization with 2 years of experience using mobbing in another company and has been mobbing with the current organization for another 18 months.

The initial motivation to encourage mobbing with the current team was to use it as a means to up-skill members in the team in coding and testing good practices, as well as for the team to get a shared understanding of expectations of code quality. The team's code base had become fragmented with evidence of very different approaches to coding and Test-Driven Development (TDD). Mob Programming was seen as a mechanism to level out the team members' skills and get a consistent level of code quality.

The team has nine members, six Developers and one Development Lead, one Tester and one Business Analyst (BA). Apart from one new graduate, the team is quite experienced in Agile practices and coding with development experience in the range 4-18 years. When the team was formed the members were new to each other as a team.

The team are co-located and use a Kanban-style way of working. They use a physical board to track progress on a product backlog of work items. Work items are estimated as they come closer to the top of the backlog and most of the team are planning for the top 2-3 items of the backlog, with a work item typically taking 2-5 days to complete. The BA takes a longer term planning window, considering work for the next 6 months and a coming up with a well-defined timeline of work for the next 3 months. The BA also re-visits work prioritization with the Dev Lead and Product Manager every 2 weeks.

The team has a daily standup meeting at 10am every morning. At this meeting the team collectively identifies what the most important things are that need to completed for the day based on external expectations and current work in progress. Team members indicate individual preferences for how items should be worked on, whether they are best done as a mob, a pair or solo work. They then decide collectively who should work on what. Pair and Mob Programming are the most common modes of working. A Test-Driven Development (TDD) approach is taken to developer coding and testing, with the team's Quality Analyst focusing on exploratory testing.

Some other development teams in the organization have experimented with mobbing. Some found it useful and still mob regularly. Others found it didn't suit them and have not continued.

Initially management questioned whether mobbing was an efficient use of time and resources. Moving the focus from resource optimization to flow optimization and delivering outcomes proved more convincing.

## 3   THE MOBBING SETUP

The mobbing sessions initially took place in a meeting room separate from the team's usual workplace. The specific meeting room depended on availability, but each had a large monitor for use in mobbing. Team members brought their own keyboards and a single laptop for use while mobbing. Later, a dedicated area close to the team's usual workspace was set up with a dedicated mobbing machine. In this area there was a 60" screen with a central desk which everyone sat around, as well as some partitions to reduce ambient noise.

Most days the team would be involved in mobbing, typically in two-hour sessions.

During early mobbing sessions, when mobbing was still new to the team, the team followed a rigid cycle of changing typists every 10 minutes, with everyone taking turns as the typist. Later, the team were more intuitive about when to change typists, either at the request of the current typist or another developer asking for the opportunity to have the keyboard. Less experienced developers were given preference for being at the keyboard to help identify gaps in their knowledge that could then be addressed.

Generally, it was the only the developers who were involved in the mobbing sessions, and occasionally the team's Quality Analyst and Business Analyst. Early in the mobbing experience the mobbing activity was kept quite separate from other activities. As experience was gained, the mobbing became more organic, with the team switching from pair programming to mobbing and back seamlessly, for example.

During a mobbing session, occasionally, some solo work would be needed and an individual mob member or two may break away from the rest of the mob to do this work on their own laptop for a few minutes. While all solo work required a code review, via pull request, before merging with the production code, this review was not required for pair or mob programming, since a review had effectively taken place during the coding in a mob or a pair.

## 4   OBSERVED BENEFITS AND CHALLENGES

Over the past 18 months, as the team has gained experience with regular mobbing, a number of challenges have been addressed and some benefits to the team, have been realized. This section describes these challenges and benefits, together with some explanations and evidence.

### 4.1 Benefits

There were several benefits that accrued over time as a result of building mobbing into the team's way of working. These are now described.

Regular (daily) mobbing reduced the number of in-progress work items. It was natural with this way of working to finish work items before starting on others. This appeared to improve the team's productivity by changing the workflow to a "pull" approach to work items and reducing context switching.

Code ownership moved from individual ownership to the team ownership. For example, team members now refer to "our" code, rather than "my" code or "your" code".

The team members have become more consistent in their approach to coding and code design. Unlike the situation before mobbing, it is now difficult to distinguish who wrote a particular piece of code just by looking at the code. There is a consistent style and approach across the team.

Team members have become more consistent in the tools used for development and more effective in their use of the tools. This has increased productivity, with the tools fading into the background more. For example, changing from team members using their favoured text editor for coding to one everyone uses has reduced one of the overheads of changing typists during mobbing.

The developers gained a broader knowledge of the system, compared to the more specialized areas they started with. Previously there was a very clear divide in skills between the front end code base and the back end services, Now all the developers are comfortable and capable of working in both the front end and back end services and in almost any area of the system. This means work can be shared in the team more easily, design decisions are better informed, collaboration is easier, and the risk of knowledge being lost if a team member leaves is lowered.

The team had more confidence in their new code, particularly when working on complex areas of the codebase. This resulted in, for example, the team being more aggressive about releasing new code into production.Team morale has improved since the team starting mobbing regularly. Written feedback from team indicates that regular mobbing was a factor in this. In the weekly retrospectives team members regularly stated that they were enjoying mobbing and the learning it was enabling.

Onboarding a new novice team member (a graduate) was quicker than the team leader's previous experience with other similar graduates with no mobbing. For example, her learning was at an accelerated rate after a few months compared to experience with others.

Confidence in the predictability of work became higher. In contrast to mob coding, people working on their own would often get stuck on a problem for too long before asking for help, throwing the team's estimates out. The team were noticeably better at delivering as estimated with regular mobbing.

## 4.2 Challenges and risks

There were also several barriers to effective mobbing that were overcome in order to embed mobbing into the work practice, as well as a several risks that became apparent as experience was gained.

It takes some effort to get mobbing going, and of value. People need to believe that it will be valuable and is worth the extra overhead and change of mind set. There is a risk that this is not accepted by all team members. For example, initially one team member preferred to work on their own, even isolating themselves with earphones, which added to the effort of instigating mobbing.

Initially code generation was slower using a mobbing approach. For example, if someone in the mob had limited understanding of the area being worked on, the learning overhead would slow the entire team's work pace. Over time this has become less of a problem and in fact contributes to benefit 5., but there is a risk this can reduce momentum with mobbing if time pressure becomes the dominant short-term motivation.

Interpersonal interactions are more frequent and intensified with mobbing, this can impact the group's ability to deliver work. Interpersonal challenges need to be resolved quickly or they become an insurmountable barrier to mobbing.

Related to the previous risk, there is a risk that an existing interpersonal challenge between two people is amplified with regular mobbing. For example, there were two people in the team who did not work well together and avoided each other, but with mobbing there was an expectation that they would be working closely with each other, and so couldn't avoid each other. There is a risk that team members who have a preference to work on their own are more visible and may be perceived as non-team players and become isolated from the rest of the team. There is also a risk that a team member who does not have good interpersonal skills may find it difficult to communicate with the rest of the team, who may have better developed interpersonal skills. This may result in that person feeling isolated.

Finding a suitable work space and equipment was a challenge initially. Booking an available meeting room some distance from the work area and potentially a different room each time, was too disruptive and unpredictable for mobbing. Even with a dedicated work space for mobbing, the immoveable furniture and networking had to be changed to allow the team to all sit at a desk and see the large screen. In addition, it was initially difficult to get a large screen and laptop dedicated to mobbing sessions.

Initially different team members' laptops were used as the mobbing machine. The unpredictability of the features specific to that machine and the mixed familiarity with the editor used on the machine was a barrier to mobbing effectively.

It was a challenge to predict accurately if a new member of the team would sustain their motivation to mob, which could be disruptive to the rest of the team's mobbing. For example, one new team member was active and enjoying mobbing for a few weeks, but after that lost motivation, and stopped.

The role of the Quality Analyst in mobbing took a while to understand and stabilize. Initially the Quality Analyst would join the mob at inappropriate times, such as during intensive coding sessions, and would be unable to contribute. This affected the Quality Analyst's morale negatively. With more experience, it was found that the Quality Analyst was most valuable during mobbing sessions that were focused more on exploring a problem. Also, it was found to be useful when the Quality Analyst was not part of the mob, but close-by and available for ad-hoc questions and comments.

## 5 LESSONS LEARNED

This section summarises the main lessons learned from reflecting on the experiences of mob programming over the past 18 months in the case organization. This includes lessons related to team size and dynamics; elements of the mobbing process and its evolution; the mobbing work space; the type of work best suited to mobbing; and long-term productivity.

The size of the mobbing team that emerged as being most effective with the case team was mobs of 3-4 people. When forming mobs larger than 4 often people in the mob would self-exit, feeling they were not contributing much. The trade-off between increased quality and the reduced pace of larger mob teams was often perceived not worth it unless complex parts of the code base were being worked on.

Often people would come and go from a mob during a mobbing time-box, whether as a break or to do pair or individual work. Initially these comings and goings could be disruptive to the flow of the team, but after some experience and reflection, team members learned to leave and rejoin the mob quietly and discretely. Another long-term benefit of mobbing was that members of the team were able to take leave without there being any major impact on the ability of the team to continue to deliver work.

New team members need time to adjust to working in a mob. They can become overwhelmed if just assigned to a mob and expected to understand when mobbing is appropriate or not. A combination of independent and mob work, with time for reflective discussion about mobbing became a good approach for easing new team members into mobbing.

One of the aspects of the mobbing process is changing who drives. Swapping the driver had some important principles to enhance the effectiveness of mobbing. Early in the mobbing experience there was a tendency to let the "expert" drive for too long, since they made progress quickly. However, this could easily lead to the rest of the mob becoming passive spectators, while they watched the expert solve the problem. Conversely it would have been easy to avoid the novice driving at all, since they were often reluctant because of their lack of knowledge and slowed progress. However, having the novice drive, even preferentially, had a number of benefits that resulted in accelerated learning and a more diverse discussion that included a novice point of view.

Time-boxing mobbing sessions were found to be effective, with 2 hours a maximum. Individuals in the mob would take independent regular short breaks every 25-30 minutes worked which helped keep the team members fresh. Because often these breaks were taken at different times it allowed a piece of work to continuously progress towards completion which further increased the flow of work.

The need for a prescriptive process for mobbing reduced as the team members' experience with mobbing increased. As the team moved from novice to competent and expert in their mobbing their actions were less determined by rules and became more intuitive. The team need to be empowered to do this.

Setting up a suitable work space with the right equipment was an important success factor for mobbing. Some of the characteristics of an effective workspace observed are:

    1.  The mobbing workspace should be close to the non-mobbing desks and daily workspace of the team. This lowers the barrier to switching to and from a mobbing mode. It should include a desk that the entire mob can sit at with a good view of the central large screen (40-inch screen as a minimum size). Also, enough desk space is needed for some mob team members to occasionally work in solo or pair mode.

    2.  A dedicated machine for mobbing is needed. This lowers the friction of getting into mobbing mode since the machine is always available, and it can be set up with the right tools and hardware for mobbing, so there is no setup overhead each time. Prior to a dedicated machine, one of the mob's laptops would be used. If they wanted to leave the mob for solo work for a while, an alternative machine would need to be found.

    3.  The mobbing area should have some boundaries and screens to identify it as a separate area for mobbing and lower surrounding noise and visual distractions.

Some work was well suited to mobbing and some was not. For example, mobbing proved to be effective for refactoring code. Mobbing was also used when there was a need for someone in the team to share their knowledge with the team. When the team was starting new work that was complex and how to even begin was uncertain, the team would mob, switching to pairs and back as different areas of investigation were identified. Generally, if the coding involved a high level of complexity or would have a high impact if in error, the team would mob, and would not, otherwise. For example a mob would not be involved in a simple UI change. If the knowledge was already shared and complexity low, pair programming was the usual mode. Very occasionally a time pressure to deliver multiple work streams to a very tight deadline was a factor in deciding not to mob.

There were several factors related to the team's long-term productivity that were impacted by regular mobbing. This included a reduction in multi-tasking, fewer interruptions, a higher level of code craft, less technical debt, and fewer delays because of unavailable information.

Another lesson, in hindsight, is that identifying the needs of the sponsors of the team as well as the needs of the team are important. We identified the individuals in the teams needs but did not identify clearly the needs of the organization on the team. If we had, we would have probably had more support for mobbing from the start.

## 6 OTHER PRACTITIONERS: A PRELIMINARY SURVEY

Having reflected on the experience of mobbing in the case organization, we wanted to get an idea if others were implementing mobbing in a similar way to us and if they had similar experiences and perceptions of it. This section reports on an informal survey of practitioners of mobbing around the world. Questions were asked about the individual and team context; their mobbing practice; the importance of personality; and the perceived benefits. The survey was conducted online and participants were invited through word of mouth. A few mobbing practitioners known to the authors were invited to participate and invite their own contacts to participate. The survey was available online for 1 month in October 2017 and 82 respondents completed the survey in this time (available at https://myobfuturemakersacademy.typeform.com/to/rHH0rV)

### 6.1 The Respondents

Most of the respondents were Developer/Coder (72%) or Team Leaders (12%). A few Testers (5%) and Business Analysts (1%) also participated in the survey. 10% of the participants classified themselves as a role other than these.

Almost half of the participants had 1 or more years of experience mobbing (49%) with 5% over 5 years' experience. 12% of those surveyed had less than 3 months experience with mobbing, with the rest between 3 and 12 months' experience.

The personality traits of the respondents is not strongly skewed towards introversion or extroversion, with a slight bias towards introverts. Almost half the respondents (47%) considered themselves to be introverts, with 9% of them labelling themselves strong introverts. Almost one third of the participants (32%) perceived themselves as extroverts with 9% being strong extroverts (9%). 22% did not consider themselves to be particularly introverted or extroverted.

### 6.2 The Context

Typically mobbing was used by teams using an Agile software development process such as Kanban (56%) and Scrum (17%), with the rest (18%) grouped into "Other". The software development teams which the respondents belonged to typically had 4-8 team members (71%). Some teams were larger, with 9 or more members (22%) and a few had 2 or 3 team members.

### 6.3 Mobbing in Practice

The size of a typical mob was a fixed number between 3 and 5 people (81%), with half of the respondents in mobs of 4. A few mobs were larger with 3 respondents (4%) working in mobs of 7 or more. The size of the mob was unpredictable and varied frequently for 13% of the respondents. Two of those surveyed worked in a mob of two people.

Around half (51%) of the respondents felt their predominant mode of doing software development work was in a mob. This compared to one third mostly working in pairs and 16% view working alone as their usual way of working.

Just over half (51%) of those surveyed work in a mob most days with 15% mobbing at least once a week. Another 7 (9%) participants worked in a mob at least a few times a month, while 26% described their use of mobbing as sporadic.

In a day that teams did mob, two thirds (67%) of the teams mobbed for most of the day, while the other third worked in mob for only a couple of hours in a typical day.

### 6.4 Personality Traits

There have been some observations that mobbing may not be effective if the personality traits of individuals are not suited to close collaboration, particularly introverted people [7]. For this reason, perceptions of high and low impact personality traits were investigated in the survey. When asked to pick one personality trait of team members that has the *biggest* impact on being effective in working in a mob, 57% of respondents chose "Openness to new experience", 17% "Agreeableness", 13% "Conscientiousness", and 2% selected "Emotional stability". No respondent chose "Degree of extroversion" as important and 11% thought that a personality trait other than those listed has the biggest impact.

When asked the converse question, which personality trait has the least impact on effectiveness in working in a mob, the results aligned with the previous question with "Degree of extroversion" being viewed as the least impactful by 67% of the respondents and "Openness to new experience" by none of them. Between these extremes were "Conscientiousness (13%), "Emotional stability" (11%) and "Agreeableness" (9%).

### 6.5 Benefits

When asked the general question if they saw value in doing mob programming, 100% of the respondents answered in the positive.

The participants were presented with a list of five potential benefits of Mob Programming and asked to indicate which ones they thought applied to their experiences of mobbing. The benefits presented in the survey were based on those found in [6]. 89% of the respondents indicated that learning from others was a benefit. An increase in code quality and the opportunity to share with others were viewed as benefits in the experience of 79% of the participants. Team participation (73%) and quicker problem solving (51%) were also seen as benefits in mobbing. Some respondents (15%) thought there were other important benefits.

## 7 DISCUSSION

The results of the survey are discussed and compared with the related findings from the case organization in this section.

Mob Programming is most strongly associated with an Agile way of working in the survey participants, and is the context of mobbing with the case organization. This is not surprising since it embodies the Agile values of collaborative programming, and could be viewed as a natural extension of Pair Programming from eXtreme Programming practice.

The usual mob size for the case organization (3-4) aligned well with the experience of those surveyed as discussed in Section 5. The team size 3-4 also aligns with other research suggesting that the optimal team size for collaborative complex problem solving is 3-5people [5].

The regularity, frequency and amount of time spent Mob Programming with the survey participants leaned towards daily mobbing for extended periods of time, as was the situation with the case team, where it became the default mode of programming.

In the survey, the mobbing teams were balanced between introverts and extroverts and also considered this distinction as unimportant to effective mobbing. This does not support the view that introversion may be a barrier to mobbing.

For effective mobbing the survey respondents clearly considered it important that the team members were open to new experiences and agreeable. The former may relate to the experimental nature of mobbing for teams, with practice and principles still in their early stages of experience. This implies that team members need a willingness to try out mobbing, even with some uncertainty in the process and the outcomes. Agreeableness may align with the case study where it was observed that mobbing amplified any problematic relationships between mob members since this way of working assumed everyone worked closely together. Agreeableness may also imply a willingness to listen to others' ideas, accept that someone else's idea may be better than yours, and sometimes compromise your ideas.

The emphasis that the survey participants placed on learning and sharing from others as positive outcomes of mob programming aligns well with the team in the case study. This is illustrated by the new graduate in the case organization, who, after six-months of mobbing, notes she has:

*"…learned a lot from mobbing with more experienced developers, without me knowing it. I guess the results are only seen a bit later. Overall, mobbing has really helped my current team deliver, spread and solidify our skills/knowledge.*

It is informative to contrast this with her initial experience where, after only two weeks of mobbing she feels she is *"not bringing value to the team… they work too fast for me so I'm not given time to come up with a solution myself."*

The additional benefit of quicker problem solving when Mob Programming, identified strongly by survey participants, is also supported by the team in the case organization. As one developer put it: *"Mobbing was a great way to perform collaborative problem solving and context sharing".*

## 8 CONCLUSION

This paper describes the process, experiences, benefits and challenges of Mob Programming for a team in a software product-driven organization. Lesson learned over the past 18 months of Mob Programming, as well as success factors, are drawn from the experience so far. This provides some guidance for others experimenting with Mob Programming in a similar context. Overall Mob Programming in the case team has been positive, has become the usual way of working, and has shown promising potential long-term benefits. A preliminary online survey of 82 Mob Programming practitioners around the world shows general alignment with the case team's experience and perceptions.

Reflecting on the team's experience and the survey results has suggested several areas that need further investigation. In what contexts is it better to code in a mob or a pair or solo? What are the perceptions of the mob team members on the value of mobbing? Are there any theories from other disciplines underpin mobbing practice and explain and predict outcomes? Some quantitative empirical evidence of the benefits of mobbing (or otherwise) would be informative.